\documentclass[sigconf]{acmart}
\usepackage{booktabs} % For formal tables
\usepackage{float}
\usepackage{multirow}
\usepackage{graphicx}
\usepackage{array}
\usepackage{xcolor}
\usepackage{soul}
\usepackage[utf8]{inputenc}
\usepackage{subcaption}
\usepackage{xfrac}
\usepackage{listings}
\usepackage{fancybox}

\newcommand{\ignore}[1]{}
\newcolumntype{P}[1]{>{\centering\arraybackslash}p{#1}}
\newcolumntype{M}[1]{>{\centering\arraybackslash}m{#1}}
\AtBeginDocument{%
  }

\copyrightyear{2026}
\acmYear{2026}
\setcopyright{cc}
\setcctype{by-nc-nd}
\acmConference[CHI EA '26]{Extended Abstracts of the 2026 CHI Conference on Human Factors in Computing Systems}{April 13--17, 2026}{Barcelona, Spain}
\acmBooktitle{Extended Abstracts of the 2026 CHI Conference on Human Factors in Computing Systems (CHI EA '26), April 13--17, 2026, Barcelona, Spain}
\acmDOI{10.1145/3772363.3798629}
\acmISBN{979-8-4007-2281-3/2026/04}

\acmISBN{978-1-4503-XXXX-X/18/06}

\begin{document}
\setlength{\tabcolsep}{2pt}

\title{StoryComposerAI: Supporting Human-AI Story Co-Creation Through Decomposition and Linking}

\author{Shuo Niu}
\email{shniu@clarku.edu}
\orcid{0000-0002-8316-4785}
\affiliation{%
  \institution{Clark University}
  \streetaddress{950 Main Street}
  \city{Worcester}
  \state{Massachusetts}
  \country{USA}
  \postcode{01610}
}

\author{Dylan Clements}
\email{dyclements@clarku.edu}
\orcid{0009-0003-4672-1524}
\affiliation{%
  \institution{Clark University}
  \streetaddress{950 Main Street}
  \city{Worcester}
  \state{Massachusetts}
  \country{USA}
  \postcode{01610}
}

\author{Marina Margalit Nemanov}
\email{MNemanov@clarku.edu}
\orcid{0009-0008-9051-377X}
\affiliation{%
  \institution{Clark University}
  \streetaddress{950 Main Street}
  \city{Worcester}
  \state{Massachusetts}
  \country{USA}
  \postcode{01610}
}

\author{Hyungsin Kim}
\email{HyuKim@clarku.edu}
\orcid{0000-0002-3794-1686}
\affiliation{%
  \institution{Clark University}
  \streetaddress{950 Main Street}
  \city{Worcester}
  \state{Massachusetts}
  \country{USA}
  \postcode{01610}
}

\renewcommand{\shortauthors}{XXX et al.}

\begin{abstract}
GenAI's ability to produce text and images is increasingly incorporated into human-AI co-creation tasks such as storytelling and video editing. However, integrating GenAI into these tasks requires enabling users to retain control over editing individual story elements while ensuring that generated visuals remain coherent with the storyline and consistent across multiple AI-generated outputs. This work examines a paradigm of creative decomposition and linking, which allows creators to clearly communicate creative intent by prompting GenAI to tailor specific story elements, such as storylines, personas, locations, and scenes, while maintaining coherence among them. We implement and evaluate \textit{StoryComposerAI}, a system that exemplifies this paradigm for enhancing users' sense of control and content consistency in human-AI co-creation of digital stories.
\end{abstract}

\begin{CCSXML}
<ccs2012>
   <concept>
       <concept_id>10003120.10003121.10003124</concept_id>
       <concept_desc>Human-centered computing~Interaction paradigms</concept_desc>
       <concept_significance>500</concept_significance>
       </concept>
   <concept>
       <concept_id>10003120.10003121.10003129</concept_id>
       <concept_desc>Human-centered computing~Interactive systems and tools</concept_desc>
       <concept_significance>500</concept_significance>
       </concept>
 </ccs2012>
\end{CCSXML}

\ccsdesc[500]{Human-centered computing~Interaction paradigms}
\ccsdesc[500]{Human-centered computing~Interactive systems and tools}

\keywords{Human-AI Co-creation, Storytelling, Generative AI, LLM}


\maketitle

\section{Introduction}
Generative artificial intelligence (GenAI) is increasingly incorporated into creative work as a collaborative partner that supports ideation, personalization, and co-authorship. In tasks such as creating storyboards, human-AI co-creativity describes practices in which humans and AI systems engage as intentional collaborators, jointly contributing to the generation and negotiation of creative artifacts, rather than simply using AI as a media production tool \cite{im_humanai_2026}. In HCI, researchers have increasingly incorporated GenAI into script editing and the organization of visual-audio materials~\cite{barua_lotus_2025, wang_reelframer_2024, huber_b-script_2019, huh_videodiff_2025}. At the same time, recent HCI studies have observed that content creators already directly use GenAI to write video scripts and produce visual materials~\cite{anderson_making_2025, lyu_preliminary_2024}.
\par

However, in human-AI co-creativity, maintaining creators' control over GenAI and supporting their sense of autonomy is central yet challenging~\cite{rafner_agency_2025, barak-medina_designing_2025}. One critical issue that hampers creators' creative agency is the inconsistency and unpredictability of GenAI outputs~\cite{rafner_agency_2025, zhang_exploring_2025}. Creators must repeatedly communicate, negotiate, and correct GenAI outputs to align them with their creative goals, which leads to increased cognitive and interactional burden~\cite{rezwana_human-centered_2025}. This challenge is amplified in storytelling tasks, where GenAI outputs must maintain stylistic and narrative coherence across multiple scenes and frames~\cite{zhang_generative_2025}. Creators must produce multiple interdependent artifacts -- including scripts, characters, scenes, and visuals -- while ensuring coherence and consistency across these components~\cite{zhang_generative_2025}.
\par
GenAI inconsistency and decontextualization can harm creative agency and creators' trust in AI systems~\cite{martinez_generative_2025}. One reason for this inconsistency is that GenAI often produces unintended variations or random distortions~\cite{liu_beyond_2023}. Another reason is that it is cognitively challenging for creators to craft effective GenAI prompts that consistently produce outputs aligned with their creative intent across multiple generations~\cite{naqvi_catalyst_2025}. While the former remains primarily a technical challenge for AI research, the latter calls for human-centered design in AI-augmented storytelling tools -- specifically, how \textbf{GenAI tools can enhance creative agency~\cite{moruzzi_user-centered_2024} by capturing creators' creative intent, structuring it into editable components, and guiding GenAI to produce coherent and well-aligned materials.}
\par

A recent scoping review of human-AI co-creation in storytelling shows that prior work has largely focused on media generation and on understanding user interactions with AI-generated stories in education, gaming, and media production~\cite{im_humanai_2026}. Existing approaches typically do not support granular editing of GenAI prompts to tailor individual story components: TaleBrush~\cite{chung_talebrush_2022} focuses on generating story text; ID.8~\cite{antony_id8_2024} requires users to regenerate all visuals in response to changes in the storyline; and SARD links story elements through visualization but offers limited control over how GenAI creates and depicts figures or backgrounds~\cite{radwan_sard_2025}. Another line of HCI research examines GenAI for video storyboarding, including techniques that automate video production~\cite{chi_automatic_2021, chi_automatic_2020}, improve the coherence of single AI-generated artifacts~\cite{leake_generating_2020, hamna_kahani_2025}, support brainstorming or organization of script sections~\cite{choi_vivid_2024, wang_reelframer_2024, leake_chunkyedit_2024}, or help creators assess relevance between text and visuals~\cite{huh_videodiff_2025, huber_b-script_2019}. While these techniques incorporate GenAI to support storytelling, they provide limited control mechanisms to address specific needs for different story elements and to maintain \textit{consistency}, such as personas and locations, across multiple scenes. Consequently, creators' sense of agency is hindered by the need to repeatedly negotiate with GenAI to preserve overall storyline coherence and alignment with their creative intent.

\par

To address these limitations, this paper introduces \textbf{Decompose-and-Link Creative Intent (DLCI)}, a human-AI co-creative interaction paradigm for digital storytelling. DLCI supports creative agency by decomposing creative intent into structured, editable entities (e.g., personas, locations, and scenes) while explicitly linking these entities so that revisions propagate coherently across AI-generated outputs. We operationalize this paradigm through \textit{StoryComposerAI}, a GenAI-powered storyboarding tool that enables creators to develop coherent stories while retaining control over narrative consistency. We present a preliminary evaluation with five participants and discuss implications for future designs that support consistency-aware human-AI co-creation.

\section{Design Concepts}
\begin{figure*}[!h]
    \centering
    \includegraphics[width=\linewidth]{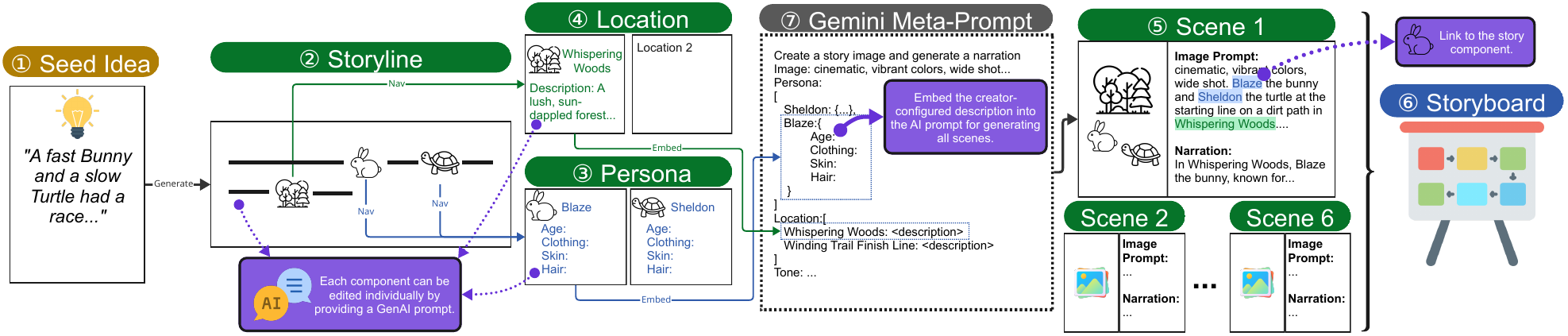}
    \caption{Decomposition and Linking (DLCI) Workflow.}
    \label{fig:design}
    \Description{A flow chart with 7 steps side by side, titled left to right as Seed Idea, Storyline, Persona, Location, Gemini Meta-Prompt, Scene 1, and Storyboard. Seed Idea contains a lightbulb and the text "A fast Bunny and a slow Turtle had a race...". This points to Storyline which contains images of a forest, a bunny, and a turtle and the text "Each component can be edited individually by providing a GenAI prompt". The forest, bunny, and turtle each point to their respective descriptions in the next two steps, Persona and Location. The location has a textual description while each persona has the fields age, clothing, skin, and hair. The next step, Gemini Meta-prompt, contains the text "Create a story image and generate a narration" with fields for an image, multiple personas, multiples locations, and tone. Arrows connect the personas and locations from the previous steps to their descriptions here. An arrow from one of the personas reads "Embed the creator-configured description into the AI prompt for generating all scenes". The next step is titled Scene 1, it contains an image with a field for an image prompt and narration. Scene 1 has the fields filled in, containing under Image Prompt "cinematic, vibrant colors, wide shot. Blaze the bunny and Sheldon the turtle at the starting line on a dirth path in Whispering Woods.." with Blaze, Sheldon, and Whispering Woods highlighted. An arrow from blaze reads "Link to the story component" with a bunny image. Under Narration the text is "In Whispering Woods, Blaze the bunny, known for his incredible speed, challenged Sheldon the turtle to a race". Below Scene 1 are an empty Scene 2 ... an empty Scene 6. The final step is Storyboard, with a image of a 3x2 storyboard on a stand.}
\end{figure*}

To operationalize the Decompose-and-Link Creative Intent (DLCI) design paradigm (\autoref{fig:design}), we propose two fundamental steps to support human-AI co-creation in storytelling: \textit{Decomposition} and \textit{Linking}. \textbf{Decomposition} focuses on identifying critical story components that meaningfully guide GenAI across creative scripts, audiovisual materials, and the final narrative. The co-creation process is structured into individually editable components, allowing creators to reflect on, provide feedback about, and revise specific GenAI-created elements of the story.
\textbf{Linking} focuses on encoding these components and their interdependencies through meta-prompting, explicitly communicating how each component contributes to the overall narrative. As a result, changes to one component propagate to related components, helping preserve the generation coherence across multi-round AI-generated outputs. 

\par
\textbf{Decomposition.}
DLCI begins with a creator-provided seed idea and enables creators to co-create with GenAI by modifying individual story components. Prior literature suggests that creating digital stories or videos involves multiple stages, including scriptwriting, segmenting the narrative into scenes, preparing visual and audio materials, and compositing these elements into a final artifact~\cite{huh_videodiff_2025}. When using GenAI to create story figures and settings, it is necessary to allow users to explicitly specify how they want to modify each story component to GenAI.
\par
Narrative structure theory~\cite{chatman_story_1978} posits that a story includes key components such as characters, settings, events, and temporal and causal structure. Following the commonly identified story components, we incorporated four critical elements: (1) seed ideas and storylines as user-defined events, (2) personas, (3) locations, and (4) scenes as representations of temporal and causal structure. The seed idea is a user-provided, high-level concept (\autoref{fig:design} \textcircled{1}). GenAI is then prompted to further develop this concept into a storyline (\autoref{fig:design} \textcircled{2}), which serves as an outline elaborated by GenAI. The persona component prompts GenAI to specify character attributes -- such as age, clothing, skin tone, and hair (\autoref{fig:design} \textcircled{3}). To establish the background, GenAI generates a location description that captures the story's setting (\autoref{fig:design} \textcircled{4}). Finally, the scene component includes an image prompt that specifies persona interactions and actions within a given location, along with a narration describing the unfolding story (\autoref{fig:design} \textcircled{5}). The scene draws on the defined personas and locations and specifies the emotional tone of each scene's narration. For the storyline, persona, location, and scene components, GenAI first generates initial descriptions and then allows users to provide feedback and modify each component individually.
\par
\textbf{Linking.}
While decomposition allows users to understand GenAI's decisions and modify individual components, the randomness and uncertainty inherent in GenAI-based visualization make scene-by-scene editing and maintaining coherence across scenes a significant challenge in human-AI co-creation. To mitigate this challenge, the \textit{Linking} mechanism employs meta-prompting to improve consistency. Meta-prompting uses structured, syntactic task descriptions that specify categories of information GenAI must incorporate during generation. In DLCI, creator-specified persona and location descriptions are automatically embedded as metadata into the prompts used to generate different scenes.
\par
In DLCI, the meta-prompting mechanism (\autoref{fig:design} \textcircled{7}) treats each defined entity as an information category and reuses the same description to generate multiple materials (e.g., \textit{Blaze} refers to a consistent description across both the storyline and individual scenes). When generating the storyline and scenes, any creator-initiated change to a persona or location automatically updates all GenAI prompts for scenes that include that entity. 
\par
DLCI supports creative agency by providing users with a greater degree of control over created components and addressing challenges identified in prior studies, which show that creators prefer to lead the creative process while mitigating unpredictability and loss of control in AI outputs~\cite{oh_i_2018}. DLCI encourages creators to explicitly communicate their decomposed intentions and exercise control over story components, while minimizing the need to repeatedly specify how each persona or location should appear across scenes. Such control over user-articulated components may enhance the sense of ownership over GenAI outputs.
% \begin{itemize}
%     \item Block-based prompting: (meta prompting, chain-of-thoughts) breakdown. GenAI performs well when you have detail instruction
%     \item Revisional Ideation, users can build an idea off a Provisional Output.
%     \item Granule change/Meta Breakdown: users can express preferences for different story component. They should be able review and modify granule elements through Prompt Decomposition.
%     \item Linking/Consistency Support. The changes need to be reflected in the overall story to keep coherence.
% \end{itemize}

\subsection{System Implementation}
\begin{figure*}[!h]
    \centering
    \includegraphics[width=1\linewidth]{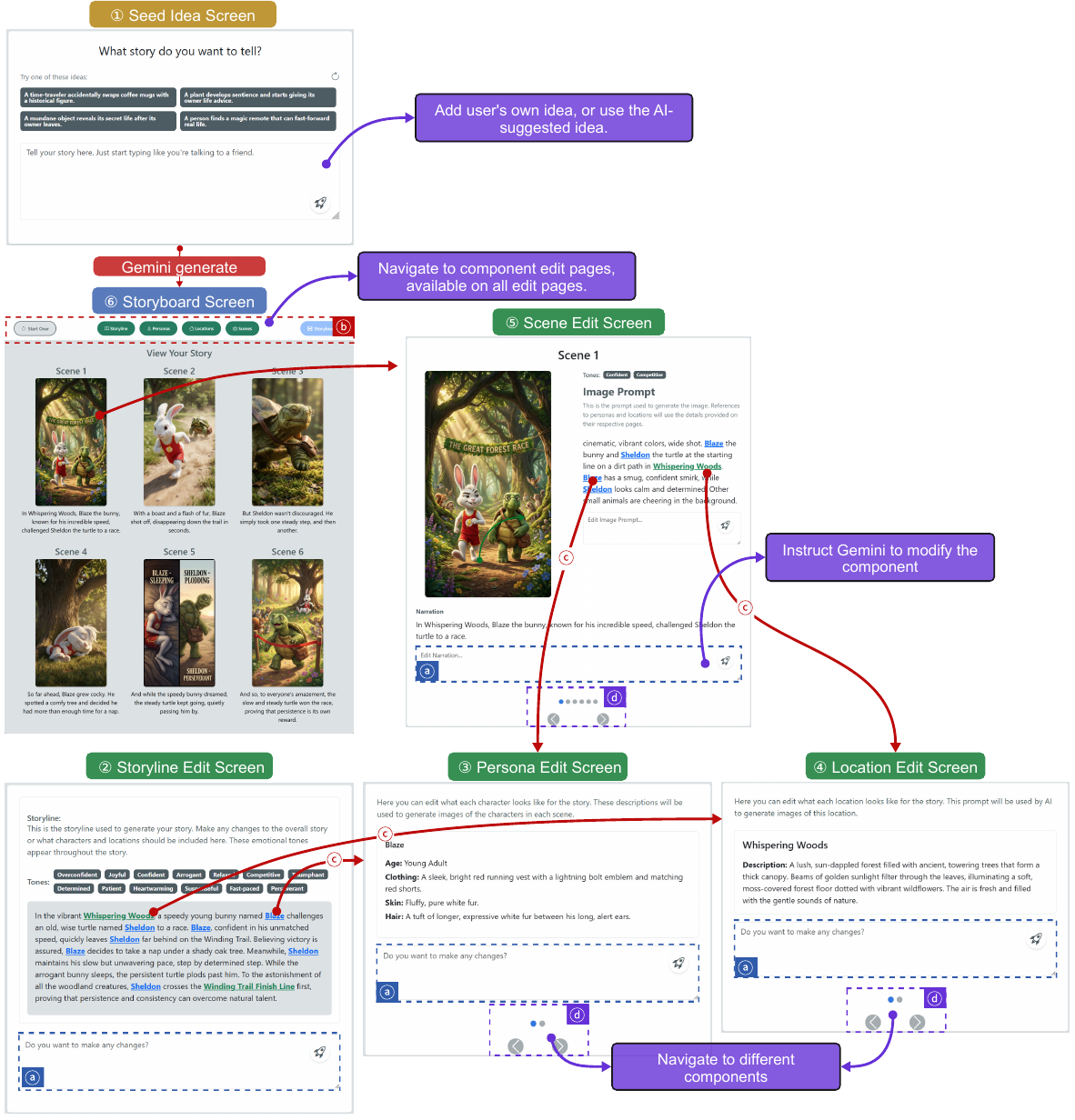}
    \caption{StoryComposerAI Interface. (1) Story creation begins with the user either selecting an AI-suggested idea or typing their own and (6) viewing the produced storyboard. From the storyboard, (b) the navigation bar allows the user to access the component edit pages, the storyboard, or start over, which is available on all edit pages as well. Interacting with an image brings the user (5) to the scene edit screen, where the user can edit either the image prompt or narration using the (a) chat box where they can instruct Gemini to modify the corresponding component. The user can use the (d) carousel to navigate to different scenes or press on a (c) highlighted persona or location name in the image prompt to navigate to either the (3) Persona Edit or (4) Location Edit Screens. By using the navigation bar on any of the edit screens the user can reach the (2) Storyline Edit Screen.}
    \label{fig:interface}
    \Description{A flow chart showing 6 screens, titled Seed idea Screen, Storyboard Screen, Storyline Edit Screen, Persona Edit Screen, Location Edit Screen, and Scene Edit Screen. Seed Idea Screen has text that asks "What story do you want to tell?", four suggested ideas in a 2x2 grid, and a chat box. An arrow from the chatbox reads "Add user's own idea, or use the AI-suggested idea". An arrow labeled "Gemini generate" leads to Storyboard Screen, where 6 images titled Scene 1 through 6 with narration under each are in a 3x2 grid. A navigation bar with 6 buttons is at the top of the screen, from left to right: Start Over, Storyline, Personas, Locations, Scenes, and Storyboard. An arrow from the navigation bar reads "Navigate to component edit pages, available on all edit pages". An arrow leads from the Scene 1 image to the Scene Edit Screen, which contains the image in the top left, the Image Prompt and Tones in the top right with a chat box, and the Narration on the bottom with a chat box. Left and right arrows are below the Narration chat box to navigate to different scenes. An arrow from the Narration chatbox reads "Instruct Gemini to modify the component". An arrow from "Blaze" points to the Persona Edit Screen, while an arrow from "Whispering Woods" points to the Location Edit Screen. These screens have the same layout, displaying the either persona or location name and description at the top, with a chat box and navigation arrows on the bottom to navigate to diffferent components. Storyline Edit Screen contains a list of Tones such as "Overconfident" and "Joyful", and a text storyline in the center describing the race between Blaze and Sheldon that starts in the Whispering Woods and ends at the Winding Trail Finish Line, with Sheldon winning. Each location and character are highlighted, pointing to the corresponding edit screen.}
\end{figure*}

To operationalize DLCI as a human-AI co-creation paradigm, we designed and implemented \textit{StoryComposerAI} (\autoref{fig:interface}), a web application that enables story creators to develop a seed idea into a complete storyboard by evaluating, modifying, and integrating different GenAI-generated story components. The system uses Gemini 2.5 Pro to generate textual content and Gemini 3 Pro Image (Nano Banana Pro) to generate story images.
\par
Users begin by providing a seed idea in the interface shown in \autoref{fig:interface} \textcircled{1}. To support brainstorming, \textit{StoryComposerAI} presents four AI-generated example ideas as illustrations. After users select an AI-generated idea or provide their own, \textit{StoryComposerAI} assigns initial values to all story components. To implement \textit{decomposition}, we use chain-of-thought prompting to guide Gemini to create story components through a structured process: (1) generating a multi-sentence storyline with 1--3 characters and 1--3 locations, (2) identifying the primary emotional tones of the story, (3) creating detailed persona descriptions with required attributes, (4) creating detailed setting descriptions for each location, and (5) generating six scenes, each with narration and an image prompt. Once generation is complete, the system saves the Gemini-generated components separately and presents the full storyboard (\autoref{fig:interface} \textcircled{6}) to allow users to assess overall coherence.
\par
The navigation bar at the top (\autoref{fig:interface} \textcircled{b}) displays the four core story components -- storyline, persona, location, and scenes -- allowing users to navigate to each component's configuration screen in any order. To support creators' negotiation with AI over each component, the storyline, scene, persona, and location edit screens each include a chat box for sending instructions to Gemini to revise the corresponding component (\autoref{fig:interface} \textcircled{a}). In the storyline edit screen (\autoref{fig:interface} \textcircled{2}), users can review the full narrative and the emotional tones applied across the storyline. The persona edit screen (\autoref{fig:interface} \textcircled{3}) presents each persona's attributes, while the location edit screen (\autoref{fig:interface} \textcircled{4}) displays detailed textual descriptions of each setting. On the scene screen (\autoref{fig:interface} \textcircled{5}), users can view the raw image-generation prompt alongside the generated image and its accompanying narration, which are later presented together in the final storyboard. Navigation buttons on the persona, location, and scene edit screens (\autoref{fig:interface} \textcircled{d}) allow users to move efficiently across components.
\par
We implement the \textit{linking} mechanism through hyperlinks across the storyline, scene, persona, and location screens. On the storyline and scene screens, clickable links (\autoref{fig:interface} \textcircled{c}) for locations (green) and personas (blue) allow users to navigate directly to the corresponding edit screens.

% \begin{itemize}
%     \item User Flow
    
%     \item Detailed Prompting
    
%     \item Full generation off one input
    
%     \item Storing element descriptions to support editing and consistency
    
%     \item Linking all appearances of names to where to edit those elements
    
% \end{itemize}

\section{Preliminary Evaluation}
To evaluate the usability of \textit{StoryComposerAI} and the underlying DLCI concept, we conducted an in-person cognitive walkthrough with college student participants from our institution (see \autoref{appx:participant}). Participants had varying levels of experience with GenAI. Each session lasted approximately 60 minutes and consisted of a 15-minute pre-study survey on demographics, a 30-minute story creation task, and a 15-minute post-study interview. Participants received a \$30 gift card as compensation. The study was approved by the Institutional Review Board (IRB) of the authors' institution.

The pre-study phase included an explanation of the task, a demonstration of the tool, informed consent, and a demographic survey. During the demonstration, the researcher walked participants through each interface page, explaining what was displayed and which story component each chat box controlled. Participants were then instructed to \textit{``create a story of your choice and edit it until you are satisfied with the final product or decide the story is complete.''} Each session employed a think-aloud protocol, during which the researcher, acting as a facilitator, operated the tool, and participants were asked to verbally articulate their intentions and desired actions. The researcher answered technical questions related to tool operation but did not provide suggestions regarding story content or inputs. 
\par
Following the creation task, we conducted a semi-structured interview informed by Lambert's digital storytelling framework~\cite{lambert_digital_2013}. Our questions examined three central factors: (1) whether DLCI enhanced participants' sense of agency over ideation, visual and narrative production, and emotional configuration; (2) whether the AI-generated content maintained consistency across scenes; and (3) whether users felt a sense of ownership over the content and were willing to share the AI-generated stories. After collecting all feedback, we first segment the answers into semantic blocks. Then three authors individually group them into themes. We then merge the similar themes into themes of design implications.

\section{Findings and Implications}

\subsection{Decomposition and Linking for Selective and Global Story Revisions}
Participants reported that decomposing personas, locations, and scenes, and allowing them to be edited through GenAI prompts, enabled targeted and selective revisions that afforded greater creative agency. Linking these components further helped maintain global consistency and stylistic coherence. For example, P5 noted, \textit{``I could change how the character looked slightly more unique [with a GenAI prompt], but without changing the soul of what was already in the image.''} P4 similarly appreciated that \textit{``you could individually change like the scene, location, personas, and the storyline without having [the entire story] muddled.''} P3 highlighted the benefits of supporting global changes: \textit{``When I want change Marble puzzle to a Rubik's cube, it was helpful cause it only regenerated the parts with the actual Marble puzzle in it.''}
\par
However, P2 identified a limitation that the current decomposition did not support adjusting narrative pacing through scene structure, stating, \textit{``In some places I felt that it used two scenes [instead of one scene]. But there was no clear way of conveying that to the system.''} 
\par
Together, these findings suggest that DLCI enables users to perform global edits while also making selective and targeted changes to local components, thereby helping to mitigate the decontextualization and inconsistency problems identified in prior research that hamper creative agency~\cite{martinez_generative_2025, naqvi_catalyst_2025}. Extending prior human-AI co-creation theory, which suggests that humans prefer to lead the task while AI provides explanations~\cite{oh_i_2018}, our findings show that users appreciate AI systems that present decomposed components, thereby supporting metacognition about what can be controlled and how such control affects overall story coherence. Beyond controllable story components, future systems should also support temporal pacing across scenes, for example by enabling creators to split or merge scenes as part of the revision process.

\subsection{Visual-First Storytelling and Negotiating Creative Intent}
Participants noted that viewing the entire storyboard with visuals, rather than only the textual storyline, was a primary driver in shaping how they believed the story creation process should be controlled. For example, P3 noted, \textit{``I can see like the whole storyboard at once. And then I can see what I don't really like and what doesn't really fit.''} P1 similarly mentioned that \textit{``[GenAI] came up with its own concept of what these people would look like, and then I can work backwards.''} Participants reported a stronger sense of agency when their intended styles were successfully depicted by GenAI. As P4 noted, \textit{``[GenAI] did add in quite a bit more than I was really expecting it to ... like the biblical style.''} Selective regeneration of styles also helped address consistency issues; for instance, P2 observed that the AI did a good job replacing a parrot with a cat and that they could instruct the system to fix certain visual errors.
\par
However, participants also identified limitations of visual-driven refinement. GenAI could not always correctly interpret users' intentions when modifying images. P3 noted that when they asked GenAI to make a Rubik's Cube demon less revealed, the system \textit{``just like removed the Rubik's cube demon from scene four, instead of putting a looming figure.''} P5 reported uncertainty about how much control they had over AI generation, noting that modifying a small part of an image took as much time as changing the entire image. P4 further observed that GenAI struggled to accurately generate non-human characters (e.g., frogs in their story) and sometimes applied decontextualized styles, such as giving a supposedly realistic persona wings.
\par
Together, these findings suggest that users prefer to negotiate the visual presentation of story components directly with AI. This contrasts with prior design patterns that use textual scripts as the primary outline and editing target~\cite{leake_chunkyedit_2024, wang_reelframer_2024, antony_id8_2024}. These results indicate that future designs should place greater emphasis on supporting users in communicating their visual intentions to GenAI and on enhancing AI systems' ability to faithfully render those intentions in visual creation.

\subsection{Creative Agency and the Tradeoff Between Efficiency and Creativity}
Participants reported that they did not rely on the system to generate ideas outright; instead, they used the tool to develop and refine ideas they already had. P2 noted that \textit{``I don't think I used the tool [for ideation]; I was able to kind of tune the story to kind of my preferences for stories.''} Similarly, P3 described their prompt as \textit{``a creative story that was like off the dome,''} and P4 used the tool to elaborate on a personal interest in \textit{``Lord of the Rings.''} While the tool provided AI-generated seed ideas, participants emphasized that meaningful creation should remain personally driven in order to preserve their sense of creative agency.
\par
However, P1 expressed a particularly negative attitude toward AI-generated content and did not feel a sense of ownership over the output, noting that \textit{``[AI] takes away humanness of art... I like that it does make it go faster often. I don't think that efficiency is worth sacrificing any of the human aspects of art or media.''}
\par
Based on these comments, we argue that GenAI systems for storytelling should enable users to lead idea generation while providing sufficient scaffolding to help elaborate their ideas. HCI designs that use GenAI for automating content production~\cite{chi_automatic_2020, chi_automatic_2021} must fully recognize and value the human need for creativity across different story components. Human-AI co-creativity theory highlights the importance of presenting actionable dimensions that users can meaningfully engage with~\cite{moruzzi_user-centered_2024}; our findings further imply that AI can enhance creative agency beyond controllable personas and locations. Future iterations of the tool may encourage creators to explicitly control and link additional key components -- such as objects or meaningful story conflicts and resolutions -- in order to better balance creative efficiency with creative agency.

\section{Future Work}
Moving forward, we aim to evaluate how DLCI influences a broader range of creative agency factors, including process control, authorship, and creative support~\cite{rhys_cox_beyond_2025, oh_i_2018}. This analysis will incorporate additional user data, such as interaction logs, satisfaction ratings, and validated agency scales. We will also examine how limitations of text-to-image models (e.g., producing varying images even when the meta-prompt remains the same) influence perceived story coherence and user satisfaction. In addition, we will investigate techniques for incorporating additional narrative components, such as story conflict and resolution, to better understand how creators control and tailor these elements~\cite{barnes_big_2012}.

\bibliographystyle{ACM-Reference-Format}
\bibliography{references}

\appendix

\section{Participant information}
\label{appx:participant}
\begin{table}[H]
    \centering
    \begin{tabular}{|p{0.05\linewidth}|p{0.06\linewidth}|p{0.12\linewidth}|p{0.5\linewidth}|p{0.23\linewidth}|}
    \hline
        ID & Age & Gender & GenAI Experience & Social Media Experience \\
    \hline
        P1 & 18-24 & Male & None & Every few months, video edits \\
    \hline
        P2 & 18-24 & Male & Daily, learning(school and personal interests), programming(syntax and generation), and information look up. Video/image generation a few times a week, for research into tool capabilities and for fun. & Photography on Instagram, 2-3 times a month \\
    \hline
        P3 & 18-24 & Male & Daily, research, school, and fun. Video/image generation a few times a month for research soon, but for fun. & None \\
    \hline
        P4 & 18-24 & Male & A few times a week as a search engine and for translation. Video/Image generation a few times a year, I've made a poster out of myself with it. & I have posted videos to youtube about the mobile game clash royale \\
    \hline
        P5 & 18-24 & Male & A few times a month for information. Video/Image generation never. & A few youtube videos, not often \\
    \hline
    \end{tabular}
    \caption{Participant Demographics}
    \label{tab:participants}
    \Description{The table summarizes demographics of five participants aged 18-24. They had various levels of GenAI experience and social media experience. One particpant had no experience with GenAI, two used GenAI tools daily, and the other two used it a few times a month.}
\end{table}

\end{document}